\documentclass[appliedmath, article, preprint, pdftex]{Definitions/mdpi}
\firstpage{1} 
\pubvolume{1}
\issuenum{1}
\articlenumber{0}
\pubyear{2023}
\copyrightyear{2023}
\datereceived{} 
\dateaccepted{} 
\datepublished{} 
\hreflink{https://doi.org/} 

\usepackage{amsmath,amsfonts,amsthm}
\usepackage{bm}
\usepackage{graphicx}
\usepackage{amssymb}
\usepackage{booktabs}

\renewcommand{\linenumbers}{} 

\Title{Optimal statistical analyses of Bell experiments}

\TitleCitation{Optimal statistical analyses of Bell experiments}

\Author{Richard D. Gill $^{1}$\orcidA{}}

\AuthorNames{Richard D. Gill}

\AuthorCitation{Gill, R.D.}

\address[1]{$^{1}$ \quad Mathematical Institute, Leiden University; gill@math.leidenuniv.nl}

\corres{Correspondence:gill@math.leidenuniv.nl}

\abstract{We show how both smaller and more reliable p-values can be computed in Bell-type experiments by using statistical deviations from no-signalling equalities to
reduce statistical noise in the estimation of Bell's $S$ or Eberhard's $J$. Further improvement is obtained by using Wilks' likelihood ratio test based on
the four tetranomially distributed vectors of counts of the four different outcome combinations, one 4-vector for each of the four setting combinations. The methodology is illustrated
by application to the loophole-free Bell experiments of 2015 and 2016 performed in Delft and Munich, at NIST, and in Vienna respectively; and also to the earlier Innsbruck experiment of Weihs et al.~(1998)
and the recent Munich experiment of Zhang et al.~(2022), which investigates use of a loophole-free Bell experiment as part of a protocol for Device Independent Quantum Key Distribution, DIQKD.\\ {\bf v4:} 1 May, 2023; \texttt{arXiv.org:2209.00702}}


\begin{document}

\section{Introduction}
In 2015 and 2016, four famous ``loophole free Bell experiments'' were performed, all of which produced statistically significant violations of Bell-CHSH (or related) inequalities.  
Their results are published in the
papers Hensen et al.~(2015)\cite{hensen}, Rosenfeld et al.~(2017)\cite{rosenfeld}, Giustina et al.~(2015)\cite{giustina} and Shalm et al.~(2015)\cite{shalm}: the Delft, Munich, Vienna and NIST experiments respectively.  The first two and the second two are strongly related. The last two, Vienna and NIST, both actually used the Eberhard inequality. The experiments in Delft and Munich had another special feature, the use of entanglement swapping for heralded entanglement generation, which we will mention later.

All four experiments have been criticised on various grounds, especially concerning imperfect randomised choice of settings, and drift of experimental parameters over time. The experimenters were themselves aware of these issues and used martingale based tests, instead of the traditional ones, to neutralise some of the problems. Anyway, later experiments have rectified many claimed defects. Nowadays a loophole-free Bell test is part of the standard methodology for device independent quantum key distribution (DIQKD). Naturally, such more complex experiments have many more sore points for sceptics to point their fingers at; this is clearly just another episode in a never-ending story. In this paper we do not enter into any of these directions. Rather, we will make the working assumption that each of the four experiments was performed in a close enough to ideal way  that, for each sub-experiment corresponding to one of the four possible setting pairs, we have data which we may think of as being made up of independent and identically distributed pairs of outcomes. We assume that the experiment satisfies the (surface level) no-signalling property, namely that the probabilities of the two possible outcomes in each wing of the experiment, given the two settings applied in both wings, only depend on the setting in the wing under consideration.

This defines a simple statistical problem in which we have just four independent multinomial distributed observations (each of just four categories), where four linear constraints (non signalling) are known to hold on the four sets of four probabilities parametrizing the four 4-nomially distributed observations. Moreover, we wish to test a null-hypothesis of a further 8 linear constraints: the hypothesis of local realism is by Fine's theorem equivalent to satisfaction of the 8 one-sided Bell-CHSH inequalities.  Now, had the constraints been linear constraints on the logarithms of the probabilities, statistical estimation and testing would have been computationally easy. But linear constraints on the probabilities themselves force us to do more work, and put us in a non-standard situation. Asymptotically optimal estimates and asymptotically optimal tests of hypotheses cannot be written down in closed-form expressions, but numerical optimisation turns out to be quite easy. A tricky point is that if the experiment is a good one, estimates of the parameters assuming the null-hypothesis of local realism to be true will usually lie on the boundary of the parameter space. Wilks' statistic (twice the difference of the maximized log likelihoods) will not have the standard asymptotic null-hypothesis distribution. Instead of the chi-square (1) distribution, we will have, if the truth is indeed on the boundary, a 50-50 mixture of chi-square (1) and chi-square (0), because at such points, asymptotically, the optimal estimate of Bell's $S$ without assuming local realism would half the time be larger than 2, half the time smaller than 2.

In this paper we explore the relation between the various possible tests of local realism, showing that in principle, under the standard assumptions, much better $p$-values could have been obtained in all four experiments with little extra computational effort. We will solve the computational issues, or at least, avoid them, using modern statistical methodology which physicists generally are not aware of. The p-values obtained with the Wilks' test turn out to be smaller than those obtained by comparing an estimate to an estimated standard error, and there are good reasons to believe that they are actually more accurate, too.

The main point is that statistical deviations from no-signalling equalities are statistically correlated with statistical variation around the theoretical values of the Bell-CHSH statistic $S$ or the Eberhard statistic $J$. Hence one can improve the observed ``naive'' values of $S$ or $J$ by subtracting a prediction of the statistical error in the observed value, based on the {\em observed} statistical deviations from the four no-signalling equalities. This is not difficult to do, and moreover leads on to other ways to improve the accuracy of the statistical estimates and tests.

We go on to look at one older and one much newer experiment: those of Weihs et al.~(1998)\cite{weihs} and of Zhang et al.~(2022)\cite{zhang}. We show that control of randomisation of setting choices has reached an unparalleled perfection in the latest experiment. This makes statistical violations of no-signalling at the manifest level (correlation between Alice's setting and Bob's outcome) a thing of the past. We suggest that it was a spurious correlation caused by the hidden confounder ``time''. The physical parameters both of random setting generators and of source and transmission lines and of detectors can drift in time. By proper randomisation of the settings, one is protected against drifts and jumps and correlation over time in the rest of the experiment. Instead of relying on assumptions which are unlikely to be true, we can design experiments whose statistical assumptions are guaranteed by the experimenter's own procedures.

\section{Background: the physics story}

For those not familiar with the physics background, here’s a potted history of Bell’s theorem and the notable experiments on what is now called ``quantum non-locality'' which led to the 2022 Nobel prize in physics for John Clauser, Alain Aspect and Anton Zeilinger. John Bell himself, the star of the story, unfortunately died quite young and unexpectedly in 1990. The main purpose of this section is to get a number of key papers into the bibliography in order to help the reader who is blissfully ignorant of this backstory but would like to know more to orient themselves. The author has written one survey paper on statistical issues in Bell experiments, Gill (2014)\cite{gill}, written shortly before the miraculous year 2015 of the first successfull loophole-free Bell experiment, immediately followed by three more. I do not go into any nitty gritty of the quantum mechanics (QM) framework of states, observables, measurements, time evolution. The point to remember is that QM does not explain what actually happens when quantum systems are measured. It only tells the physicist what the statistics will be of repeating the same preparation and measurement many times. Since the birth of quantum mechanics this has been a deep cause of discomfort, mystery and debate; still today, some physicists still search for an underlying theory of more classical nature which would actually explain the randomness in observations of quantum systems as merely the reflection of deterministic processes with initial conditions which cannot be controlled in any way. One path to the unification of relativity theory and quantum mechanics would be a description of quantum mechanics as a collection of emergent phenomena arising from a deeper hidden level where more classical physical rules are followed. Others believe that determinism must give way and relativity theory will need adjustment. Many other standpoints are possible.

The story may start with the paper Einstein, Podolsky and Rosen (1935)\cite{epr}, in which it was argued that quantum mechanics was either wrong, or incomplete. A thought experiment involving measurement of either position or of momentum of two particles in a so-called singlet state showed that each particle possessed definite values of both properties, while according to quantum mechanics, a particle only got a definite value of either property, after it was measured in an appropriate measurement set-up. The assumption in the EPR argument was a locality assumption: measuring one particle couldn’t have any influence on another, distant, particle. Discussion of the foundations of quantum mechanics subsided, under the influence of its enormous success and Feynman’s famous dictum ``shut up and calculate’’. A few stubborn individuals did continue to think and to question. David Bohm converted the EPR thought experiment into an experiment concerning the spin of two entangled spin-half particles; one can measure spin of such particles in any chosen direction but the outcome of the measurement is binary: the particle as it were chooses either the direction set by the experimenter or the opposite direction. Then came Bell’s famous (1964)\cite{bell} paper, taking the EPR-B model and now adding a new twist: instead of only the same two possible measurements on each particle, he considered several different possible measurements on each. EPR had concluded, assuming local realism, that QM is either wrong or incomplete. Bell’s conclusion was the more shocking: QM is either wrong or non local. 

Bell’s thought experiment was in 1964 far from being experimentally feasible. A few wild spirits became interested and started working towards experimental testing. Their actual expectation was that quantum entanglement would rapidly decay as particles moved further apart. They did not expect to see the signature of quantum entanglement in the measurements of particles widely separated in space. In order to get closer to experimental test, the EPR-B model was transposed from the spin of spin-half particles to the polarization of photons (each offered the choice between two perpendicular 2D orientations, instead of the choice between two opposite 2D directions). Also Bell’s original inequality was generalised to what is now called the CHSH inequality, after Clauser, Horne, Shimony and Holt (1969)\cite{chsh}. A first experiment in which violation of Bell inequalities was observed was performed by Freedman and Clauser (1972)\cite{freedmanclauser}. A big defect was that settings of polarizers was kept fixed for many consecutive photon pairs. Thus each photon had plenty of time to know how both were going to be measured. In a now world famous experiment,  Aspect et al.~(1982)\cite{aspect} managed to observe a statistically significant violation of Bell-CHSH inequalities while measurement settings were chosen while the photons were in flight. Further experiments made further refinements. For a while the most impressive was Weihs et al.~(1998)\cite{weihs}. However a big defect in all these experiments was what is called the detection loophole. In Weihs’ experiment, it appeared that only 1 in 20 photons made it from source to detection, so only 1 in 400 emitted photon pairs resulted in a measurements of both their polarizations. Already, Pearle (1970)\cite{pearle} had shown that quantum correlations could be faked by a classical and local mechanism if enough particles did not show up at the detectors, and Garg and Mermin (1987)\cite{gargmermin} showed that the critical detection rate is 83\% in an Aspect or Weihs type experiment with maximally entangled photon polarizations. That's a very long way to go from 5\%.

The subsequent decades were spent on working towards so called loophole free Bell experiments in which statistics were obtained, predicted by QM, and impossible to be explained by a classical physical mechanism without recourse to superluminal messaging or even more outlandish explanations. The big breakthrough came in 2015 with four experiments carried out in Delft, Munich, Vienna and at NIST (Boulder, Colorado). The Delft and Munich experiments used a novel technology called entanglement swapping, developed by Zeilinger and others in preceding decennia. These experiments had no ``no shows’’ at all, but rather small sample sizes. The Vienna and NIST experiments used an alternative to Bell’s inequality called the Eberhard (1993)\cite{eberhard} inequality,  which used the clever device of measuring polarization in one orientation only, merging all ``no shows’’ with the perpendicular polarization outcome, thus resulting in guaranteed binary outcomes. Eberhard had discovered paradoxically that using less than maximally entangled photons one could get away with a 67\% detection rate, so detector efficiency need not be got so high as for the old-style (Aspect, Weihs) experiments. 

The 2015 experiments were not perfect and various defects need to be honed away, but the net impact of four resounding confirmations of Bell’s genial discovery was enough for the 2022 Nobel prize committee. Research continues on using an embedded loophole-free Bell experiment as part of a protocol for creating shared secret random keys at two distant locations while communicating over public communication channels. The most promising technology is that based on the Delft and Munich experiments. Here, two distant ``solid state’’ stationary qubits are brought into quantum entanglement by having both emit a photon which meet one another and interfere at a third intermediate location, where a third collaborator Charlie measures the two photons after they have interfered, and reports his findings to Alice and Bob, who at the same time were measuring their qubit in one of several ways. Alice and Bob study the statistics of the measurement outcomes and settings corresponding just to those occasions when Charlie got a certain measurement outcome. Just as one can generate statistical dependence between originally independent random variables X and Y by conditioning on a function of X and Y, it is also possible to generate quantum entanglement between quantum systems which have never physically interacted with one another by conditioning on a measurement outcome on two emitted particles which have interacted at a third location. 

A complicated protocol now in principle allows them either to determine that there has been no interference in their communications and to distil some number of secret shared random bits, or to detect interference or imperfection and abort the process. Input into all these experiments consists preferably of independent, local, completely random, setting choices. In many experiments physical random number generators have been used, whose properties tend to slowly drift as time goes by (the experiment might last several days), and occasionally moreover jump when shocks occur (e.g., a lorry crosses the campus). At the same time, the same external processes are causing drifts and jumps in the physics of source, transmission lines, detectors. The result can be a spurious correlation between, for instance, Alice’s settings and Bob’s outcome, even though Alice’s setting could not have reached Bob’s apparatus in time to influence the measurement outcome. Both are influenced by a hidden confounder: time. Plenty of techniques are available to discount a certain amount of deviation from complete randomness in setting choices, but this leads to less transparent results, depending moreover on assumed limits on the amount of bias.

\section{The statistical model}

In a standard ideal Bell experiment, two separated experimenters, Alice and Bob, each repeatedly insert a binary setting into some apparatus and a short time after observe a binary outcome. Alice and Bob work in a carefully synchronised way, such that each setting of Alice could not reach Bob's lab before Bob's outcome was registered even if travelling at the speed of light, and vice-versa. Alice and Bob might be inserting settings which were somehow generated ``on demand''  by some auxiliary randomisation procedure, whether physical or algorithmic. Alternatively, they might be reading off settings one at a time from a pre-generated data-base. All of these possibilities have advantages and disadvantages which we do not discuss here. We will use the word ``trial'' to denote one set of four binary outcomes, namely a setting for each of Alice and Bob, and an outcome for each of Alice and Bob. In first instance, the experiment generates an $N\times 4$ spreadsheet of settings $a$, $b$ taking values, say, in the set $\{1, 2\}$ and outcomes $x$, $y$, taking values, say in the set $\{-1, +1\}$. 

Given the pair of settings $(a, b)$ used in just one trial, we will consider the pair of outcomes as being the realisations of two Rademacher (i.e., $\pm 1$ valued) random variables with a joint probability distribution which depends only on $(a, b)$; and given all the settings, we consider all the pairs of Rademachers as being independent of one another. By sufficiency, we may reduce the data to the sixteen counts $N(x, y \mid a, b)$ of trials with outcomes $(x, y)$ and settings $(a, b)$. Grouping these according to the settings, we have four realisations of four independent tetranomially distributed random vectors $(N(x, y \mid a, b): x, y \in \{-1, +1\})$, for $(a, b)$ in $\{1, 2\}\times \{1, 2\}$. By definition of the multinomial distribution and by conditioning on the settings of all the trials, the sums $n_{ab} = N_{ab} :=\sum_{x, y} N(x, y \mid a, b)$ are fixed. For each $(a, b)$, the probability distribution of the 4-vector $(N(x, y \mid a, b): x, y \in \{-1, +1\})$ is the multinomial distribution with number of cells $ = 4$, number of trials $= n_{ab}$, and multinomial probabilities $\boldsymbol p_{ab}= (p(xy \mid ab), x, y = -1, +1)$; a vector of four probabilities adding to one.

We do expect a number of constraints to hold on the 16 probabilities $p(xy \mid a, b)$. Obviously, they add up to +1 in groups of four. These constraints are called the normalisation constraints. Less obviously, we have the \emph{no-signalling} constraints. For a well conducted experiment, we believe and we will moreover assume that given all settings, the marginal probabilty distribution of Alice's outcomes doesn't depend on Bob's setting, and vice versa. Using a ``+'' to denote addition over all values of a given argument, we assume that $p_a(x) := p(x, + \mid a ,b ) = p(x, + \mid a, b')$ for all $a$ and all $b \ne b'$, and $q_b(y):= p(+, y \mid a ,b ) = p(+, y \mid a', b )$ for all $b$ and all $a \ne a'$. These equations are the so-called no-signaling equalities. A little thought shows that because of the normalisation contraints (probabilities add up to +1) and the no-signalling constraints, our 16 probabilities $p(x, y\mid a, b)$ depend on just 8 free parameters. We can take them as the four marginal probabilities of the outcome $+1$ given the local setting on each side of the experiment $p_1(+1), p_2(+1), q_1(+1), q_2(+1)$, and the four correlations  $$\rho_{ab} := p(+1, +1\mid a, b) +  p(-1, -1\mid a, b) - p(+1, -1\mid a, b) - p(-1, +1\mid a, b).$$
To be specific
$$p(x, y\mid a, b) = {\textstyle \frac14} +  {\textstyle \frac12}(p_a(x) - {\textstyle \frac12}) +  {\textstyle \frac12}(q_b(y) - {\textstyle \frac12}) \pm \frac{\rho_{ab}} 4$$
where the `$\pm$' sign is `$+$' if $a = b$ and `$-$' if $a \ne b$, and $p_a(-1) := 1 - p_a(+1$), $q_b(-1) := 1 - q_b(+1$). The eight parameters vary freely in the sense that they vary in an eight-dimensional closed convex polytope with non-empty interior, bounded by the hyperplanes determined by the non-negativity of the $p(x, y \mid a, b)$. Another way to say this, is that the vector of all 16 probabilities $p(x, y\mid a, b)$ lies in a closed convex polytope in an 8-dimensional affine subspace of $\mathbb R^{16}$, with non-empty relative interior. It is called the no-signalling polytope.

As is well known, according to quantum mechanics, the possibilities are limited to a strictly smaller closed convex subset called the quantum body, and according to local realism, they are limited even further to a polytope called the local realism polytope. The two smaller sets are both full, having a non-empty relative interior relative to the 8-dimensional affine subspace in which all are constrained to lie. For instance, the point where all sixteen probabilties equal $\frac 14$ lies in all of their relative interiors.

Each of these three convex sets are the convex hull of their boundary, and the boundaries of the two polytopes are finite sets of points. 

This has defined a nice statistical model for four independent tetranomially distributed random vectors, the only unusual feature (relative to standard statistical theory) is the no-signalling constraints on the mean vectors of the four observations. A further non-standard feature is that we are interested in testing the null hypothesis of local realism against the alternative of quantum mechanics. We have non-standard estimation problems and a non-standard testing problem.

\section{The methodology}

As we have seen, a standard Bell-type experiment with
\begin{itemize}
  \item {two parties,}
  \item  {two measurement settings per party,}
  \item  {two possible outcomes per measurement setting per party,}   
\end{itemize}
generates a vector of $16 = 4 \times 4$ numbers of outcome combinations per setting combination. 

As is now well known, this can be applied to two-channel experiments without a detection loophole,
but also to one-channel experiments and (equivalently) to two-channel experiments with $-1$ and ``no-detection'' combined,
as long as the experimental units are ``time-slots''.
The four sets of four counts can be thought of as four observations each of a multinomially distributed vector over four categories. This
probability distribution is also known as the tetranomial distribution.

We will rewrite what we discussed in the previous section in a different notation, more convenient for converting formulas into programming code in the language {\em R},
or any other modern programming language.
Write $X_{ij}$ for the number of times outcome combination $j$ was observed, when setting combination $i$ was in force. 
Let $n_i$ be the total number of trials with the $i$th setting combination. 
The four random vectors $\vec X_i = (X_{i1}, X_{i2}, X_{i3}, X_{i4})$,  
$i = 1, 2, 3, 4$, 
are independent each with a Multinomial$(n_i; \vec p_i)$ distribution, 
where $\vec p_i =(p_{i1}, p_{i2}, p_{i3}, p_{i4})$.

The 16 probabilities $p_{ij}$ can be estimated by relative frequencies $\widehat p_{ij} = X_{ij}/n_i$ which have the following variances and covariances:
$$\mathrm{var}(\widehat p_{ij}) =  p_{ij} (1 - p_{ij}) / n_i,$$
$$\mathrm{cov}(\widehat p_{ij}, \widehat p_{ij'}) =  - p_{ij} p_{ij'} / n_i\quad\text{for $j \ne j'$},$$
$$\mathrm{cov}(\widehat p_{ij}, \widehat p_{i'j'}) = 0\quad\text{for $i \ne i'$}.$$
The variances and covariances can be arranged in a 
$16 \times 16$ block diagonal matrix $\Sigma$ of four $4 \times 4$ diagonal blocks of non-zero elements. 

Arrange the 16 estimated probabilities and their true values correspondingly in (column) vectors of length 16. 
I will denote these simply by $\widehat p$ and $p$ respectively. 
We have $\mathrm{E}(\widehat p) = p \in \mathbb R^{16}$ and $\mathrm{cov}(\widehat p) = \Sigma \in \mathbb R^{16\times 16}$.

We are interested in the value of one particular linear combination of the $p_{ij}$, let us denote it by $\theta = a^\top p$. 
The vector $a$ might specify the CHSH quantity $S$, or Eberhard's $J$.
We know that four other particular linear combinations are identically equal to zero: the so-called no-signalling conditions. 
This can be expressed as $B^\top p = 0$ where the $16\times 4$ matrix $B$ contains, as its four columns, the coefficients of the four linear combinations. 
We can sensibly estimate $\theta$ by $\widehat\theta = a^\top \widehat p - c^\top B^\top \widehat p$ where $c$ is any vector of dimension 4. 
For whatever choice we make, $\mathrm E \widehat \theta = \theta$. 
We propose to choose $c$ so as to minimise the variance of the estimator. 
This minimization problem is an elementary problem from statistics and linear algebra (``least squares''). 
Define
$$\mathrm{var}(a^\top\widehat p) = a^\top\Sigma a = \Sigma_{aa},$$
$$\mathrm{cov}(a^\top\widehat p, B^\top \widehat p) = a^\top\Sigma B = \Sigma_{aB},$$
$$\mathrm{var}(B^\top\widehat p) = B^\top\Sigma B = \Sigma_{BB};$$
then the optimal choice for $c$ is
$$c_{\text{opt}} = \Sigma_{aB}\Sigma_{BB}^{-1}$$
leading to the optimal variance
$$\Sigma_{aa} - \Sigma_{aB}\Sigma_{BB}^{-1} \Sigma_{Ba}.$$

In the experimental situation we do not know $p$ in advance, hence also do not know $\Sigma$ in advance. However we can estimate it in the obvious way (``plug-in'') and for $n_i\to\infty$ we will have an asymptotic normal distribution for our ``approximately best'' Bell inequality estimate, with an asymptotic variance which can be estimated by natural ``plug-in'' procedure, leading again to asymptotic confidence intervals, estimated standard errors, and so on. 
The asymptotic width of this confidence interval is the smallest possible and correspondingly the number of standard errors deviation from ``local realism'' the largest possible. 
The fact that $c$ is not known in advance does not harm these results.

The methodology is called ``generalized least squares''. 

One can go further. It is sensible to use these estimates as the starting point of Newton-Raphson iterations searching for the maximum  over the two polytopes of interest of the multinomial log likelihood instead of the quadratic loss function. Subsequently we may compute the Wilk's generalised log likelihood ratio test, evaluated through its asymptotic chi-square distribution (actually, because of boundary issues, a mixture of chi-square distributions with different numbers of degrees of freedom). In fact, as far as asymptotic results are concerned, just a single Newton-Raphson iteration should produce asymptotically optimal estimates.  Switching from generalised least squares to minimize a variance to one-step Newton-Raphson on the log likelihood maximum likelihood and then to true maximum likelihood typically gives a better approximation, at each step, to the asymptotic distribution. Asymptotically all three are equivalent. We will investigate what happens when we indeed try to obtain more reliable estimates and tests in this way.

\section{The results}

We first performed these computations on the data sets of the four famous loophole-free Bell tests of 2015. The analyses were performed by scripts written in the {\em R} language \citep{R}, and published on the website \url{https://rpubs.com/gill1109} using the IDE {\em RStudio}. For each of the four experiments, we did one analysis computing the CHSH quantity $S$ and one computing Eberhard's $J$; we computed standard errors and p-values using the asymptotic normality of the multinomial distribution; then we computed optimized version of $S$ and $J$ by subtracting the linear combination of the four observed deviations from the no-signalling equalities, which maximally reduces the (estimated) variance, thus leading to a maximal p-value. As must be the case, the optimised $S$ and $J$ are related by the theoretical identify $S = 2 + 4 J$, and the resulting standard errors differ by a factor of $4$; the p-values based on asymptotic normality are identical. The p-values are all approximate, being based on large sample multivariate normal approximations to the distribution of the 16 raw counts, and estimated covariance matrices.

After that, we attempt to estimate $S$ and $J$ by maximum likelihood for four independent tetranomially distributed vectors of counts; tests of the hypotheses of interest are now computed using the generalised likelihood ratio test and its asymptotic distribution: a 50-50 mixture of chi-squared (1) and chi-squared (0) distributions. Thus we avoid the initial step of approximation of multinomials by multivariate normals. In the case of NIST and Vienna, this leads to numerical problems: the numbers of trials are so large that numerical optimisation starting at the earlier obtained estimates terminates immediately. From the numerical point of view, the earlier obtained estimates are already indistinguishable from maximum likelihood estimates. The numerical problem is in fact a non-problem.
Indeed, the sample sizes are so large in these two experiments that systematic violation of model assumptions is much more important than statistical variation of observed counts. Our assumptions of constant physical parameters throughout the whole run are possibly wrong. Time variation in the physics, and in particular, time drifts in the physical random number generation of settings, means that to a small extent ``no signalling'' is violated: from observing her local statistics, Alice could in principle, to a tiny extent, guess Bob's settings better than assuming independence and constant probabilities. This can to a large extent be taken care of by using martingale based tests based on assumptions about the randomnness of the settings, instead of tests based on assumptions that trials (under the same pair of settings) are i.i.d. We will return to that option at the end of the paper when we discuss the Zhang et al.~(2022) experiment.

We published the results of running 8 {\em R} scripts back in 2019.  They can be found on the following web pages:

\url{https://rpubs.com/gill1109/OptimisedDelft_2}

\url{https://rpubs.com/gill1109/OptimizedMunich}

\url{https://rpubs.com/gill1109/OptimizedNIST}

\url{https://rpubs.com/gill1109/OptimizedVienna}

\url{https://rpubs.com/gill1109/AdvancedDelft}

\url{https://rpubs.com/gill1109/AdvancedMunich}

\url{https://rpubs.com/gill1109/AdvancedNIST}

\url{https://rpubs.com/gill1109/AdvancedVienna}

The addition of ``underscore 2'' means that an original document from 2019 has been improved in 2022 by some minor editing. We will next discuss our findings for the Delft experiment at some length, and then point out any notable features of the analyses of the other three experiments. 

We also added similar analyses of the Weihs et al.\ (1998)\cite{weihs} experiment in Innsbruck and of the Zhang et al.\  (2022)cite{zhang}: experiment (on DIQKD) in Munich: 

\url{https://rpubs.com/gill1109/OptimizedWeihs}

\url{https://rpubs.com/gill1109/AdvancedWeihs}

\url{https://rpubs.com/gill1109/OptimizedDIQKD}

\url{https://rpubs.com/gill1109/AdvancedDIQKD}

\section{The experiments}

\subsection{Delft}

We proceed to read off some of the statistical results of our analysis\\
\noindent \url{https://rpubs.com/gill1109/OptimisedDelft_2} of the Delft experiment. The observed frequencies (counts) and relative frequencies are also reproduced in the appendix of this paper.  One observes $S = 2.4225$ with an estimated standard error of $0.2038266$, giving a z-value of $2.07284$ and an approximate p-value of $0.0190936$. In round numbers, $S=2.42(0.21)$ giving us a p-value of $0.02$. We can slightly reduce the standard error and the p-value by optimally subtracting noise, by assuming that in reality no-signalling is true (the probabilities of Alice's outcomes do not depend on Bob's setting and vice-versa). The relative frequencies do vary slightly with the setting on the other side. This is (under our model assumptions) pure noise, but it is pure noise which is correlated with the noise in the estimate of $S$. The slightly better estimate (in the sense of lower variance) is $S = 2.462658$ or in round numbers $S = 2.46(0.20)$ and the p-value is now $0.01$. 

All p-values here are approximate: they assume that the asymptotic normal distributions of the statistics gives a good approximation to the actual distribution, and the asymptotic theory is conditional on our assumption of four independent tetranomially distributed observed count vectors.

Lovers of the Eberhard test might prefer to look at Eberhard's $J$ which takes the value $0.1195162$. Under our assumptions $E(S) = 2 + 4 E(J)$, so the observed values of $J$ and $S$ do correspond nicely. However, for this data, $J$ has a much higher estimated variance than $S/4$. Its estimated standard error is $0.09475703$ leading to a z-value of just larger than 1 and a p-value of about $10\%$, terrible compared to that of $S$, namely, about $1\%$. When we improve $J$ by the same noise reduction strategy as we applied to $S$, we end up with an improved estimate of $J$ exactly equal to the improved estimate of $S$, divided by 4; and the same p-value.

In all cases, the experiments all exhibit violation of the one-sided Bell-CHSH inequality in which three correlations are added and one is subtracted. By recoding of outcome labels if necessary, we have arranged that the exceptional correlation (large, negative) corresponds to setting pair (2, 2); the other three correlations are large and positive.

Next we take a look at the script 

\noindent \url{https://rpubs.com/gill1109/AdvancedDelft}. Here we stick to multinomial distributions. We assume no-signalling and estimate the 16 probabilities corresponding to the 16 relative frequencies by maximizing the likelihood (a) without any restriction, (b) assuming local realism. 

The two numerical optimizations give no problems. We take as initial estimates, the estimates we obtained before, using approximate normality. Without assuming local realism, there are 8 free parameters: each of the four tables has its own correlation; then there are the marginal probabilities of Alice's outcome $+$ under each of Alice's settings, and similarly for Bob. We estimate $S$ as the sum of three of the correlations minus the fourth; $J$ is estimated by $(S-2)/4$. Under local realism, the maximum likelihood estimate of $S$ (corresponding to adding the first three correlations and subtracting the fourth) exactly equals 2: the data quite strongly violates the corresponding one-sided CHSH inequalitty. The fourth correlation is an affine function of the other three. We have to optimize over 7 free parameters, forcing $S = 2$.

We test the hypothesis of local realism by comparing the maximised log likelihood in the two situations. Large sample theory (Wilks' test) tells us that under the null hypothesis, twice the difference should have a mixture of 50-50 a chi-square distribution with one degree of freedom, and 50-50 be identically equal to zero. The latter case occurs when the unconstrained estimate of the 8 parameters is a point inside the null-hypothesis. This means that in our case, our p-value according to the Wilks test is $0.04704162/2 = 0.02352081$, or in round numbers, a p-value of $2.4\%$. 

This seems less attractive than the previously obtained $1\%$; however, experience shows that the p-value obtained from the log likelihood ratio based Wilks test is a better asymptotic approximation than the p-value based on a z-value based on the approximate normal distribution of an estimator. Certainly, a shorter chain of approximations are being made when we stick closer to the underlying multinomial distributions. In the asymptotic theory, the asymptotic chi-square (1)  distribution does correspond to squaring an asymptotically standard normally distributed quantity, so everything is still based on the central limit theorem and the law of large numbers. Still, there are sound theoretical explanations for the afore-mentioned practical experience, based on higher order asymptotic theory, where one also looks at the rate of convergence to asymptotic normal distributions. Essentially, the Wilk's approach eliminates a second order term due to skewness. The statistical approach automatically eliminates a possible asymmetry leading to a nonzero coefficient of skewness in the estimators. The Wilks' approach is invariant under reparametrization; whereas many unfortunate parametrizations lead to skewness.

\subsection{Munich}

In this experiment, the observed value of $S$ was the impressive $2.609047$, with estimated standard error $0.2484456$. In round numbers, $S = 2.61(0.25)$. This resulted in a very nice p-value of $0.007114475$, or in a round number, $0.7\%$. The optimised value of $S$ was $2.582261$, so slightly less. As must be the case, its estimated standard error was slightly smaller too, resulting in much the same p-value $0.008782296$, which one should report in a round number as $0.9\%$

Maximum likelihood estimation based on the multinomial likelihood worked without a hitch. The Wilks test gave a p-value of $0.04104834/2$ which we can fairly report as $2\%$. Again, theory and experience suggest this is a rather more reliable number than the p-values based on a z-statistic.

\subsection{NIST}

Now we find $S= 2.000092$ with a standard error of $1.572689e-05$. This gives us a z-value of $92/15.7 = 5.859873$. The p-value is astronomically small $2.062969e-09$. Eberhard's $J$ would be preferred my many for this situation. It gives a z-value of $4.778576$ and a p-value of $8.827054e-07$. Optimizing the variance gives $S = 2.000051$, so closer to $2$, but of course also a smaller standard error, and in this case a much smaller standard error, resulting in a z-value of $7.637903$ and a p-value of $1.110193e-14$. However all these p-values need to be taken with a very large pinch of salt, since the convergence to asymptotic normality is worse and worse, in terms of relative error in true and approximated tail values, the further in the tail we are. But certainly, it would have been nice in the published paper to talk about $7.6$ standard errors deviation from local realism rather than $4.8$.

Maximum likelihood based on multinomial distributions resulted in warnings being issued by the numerical procedure. Essentially, the optimization of the log likelihood over 7 or 8 parameters does not succeed in moving the estimates away from values found by generalised least squares after switching to multivariate normal distributions, so a standard numerical optimisation program just gives up after a while, after uttering a lot of complaints: it cannot improve on the initial estimates by an amount greater than expected numerical accuracy. No matter. Still, the Wilks' test provides a perhaps slightly more reliable p-value than the z-tests we already discussed. The Wilks' test statistic comes out as $57.19689$ corresponding to a z-value of the square root of that number, $7.562863$;  so the message is $7.6$ standard errors, and a p-value of $1.971474e-14$.

\subsection{Vienna}

This experiment had an about 10 times larger sample size than NIST. We find $S = 2.000028$, so the difference with the local realism bound of 2 is three times smaller as for NIST. The standard error is $3.283419e-06$ so the z-value is $8.527696$. Optimizing the variance reduces it by a factor or 2 while leaving the estimate of $S$ essentially unaltered, so finally we get a square root of 2 times larger z-value of just above 12. 

Maximum likelihood  using the original multinomial has exactly the same numerical issues (or if you like -- non-issues) as in the case of NIST. It comes up with an even larger z-value than that obtained in the method based on optimising the estimate of $S$ by minimising its variance, through using the statistical deviations from no-signalling to reduce the error in $S$. In fact, we now get a z-value of $17.5$, and experience and theory says this is more reliable than the z-values mentioned so far.

\subsection{Weihs et al.}

Just for fun, we also carried out our analyses for this earlier (1998), and quite famous experiment. It was for a long time the definitive Bell type experiment, having a large enough distance between Alice and Bob that the ``locality loophole'' was closed. However, there remained a very serious ``detection loophole''. Thinking in terms of photon pairs emitted from the source, only one in 20 of the photons resulted in a detection event, and only 1 in 400 emissions of a photon pair resulted in two detections. In order to conclude that local realism has been disproved by this experiment, one must make the ``fair sampling assumption'' that detection of photons is independent of those hidden variables and independent of the settings. This experiment has $S = 2.73$ and estimating it optimally hardly changes it, resulting in $S = 2.71$. This is more than $z = 25$ estimated standard deviations away from the local realism bound $S \le 2$.
 
\subsection{Zhang et al.}

The paper \cite{zhang} is an attempt to show feasibility of using a loophole-free Bell experiment as a component in a procedure for ``device independent quantum key distribution''. Alice and Bob actually choose randomly between three settings, in such a way that some of the trials are performed with equal (or equal and opposite settings). The experiment was actually a three-party experiment similar to the earlier experiments at Delft and Munich. One studies correlations between Alice and Bob's outcomes, conditional on a particular outcome having been obtained by Charlie. This is sometimes mistaken for post-selection, but it is not. Always one studied experimental data after the experiment is completed, so if for instance one looks at the correlation between Alice and Bob's outcomes when their settings are, say, (1, 1), one is only looking at selected trials. In order for these three party experiments to be loophole-free there must be no locality loophole concerning the three parties. In particular, Charlie's outcomes must not be able to influence Alice's or Bob's measurement apparatus during the pre-allocated time-slots of the three parties. In the experiment \cite{zhang} Alice and Charlie shared a lab, in fact, they shared a lot of the electronics, so this was not actually a loophole-free experiment.

What is rather nice about this experiment, is that it seems that the random generation of settings has become very stable and close to unbiased. Our procedures for optimising CHSH led to almost no improvement in p-value.
Because of the near perfect experimental symmetry, and lack of drifts in physical parameters over time, the statistical deviations from no-signalling are hardly correlated with $S$ at all. The experiment has $S = 2.58$ with $z \approx 7$. That gives of course a tiny $p$-value of $3 \times 10^{-12}$ but this should not be taken too seriously. Since $7^2 = 49 \approx 50$ a conservative but much more reliable p-value would be $0.01$. 

Now an alternative statistical analysis, if we assume the settings are chosen again and again by independent fair coin tosses, is based on a so-called martingale test, also known as the Bell game. One says that Alice and Bob have won each separate trial (in a game they play against nature) if their outcomes are equal and neither chose setting ``2'', or their outcomes are opposite and both chose ``2''. In this experiment, the number of wins was $1357$ out of $1649$ trials, notice that $1357/1649\approx 0.82$. Under local realism (per trial, conditional on the past) the number of wins cannot have larger tail probabilities than those of  the binomial distribution with $N = 1649$ and $p = 0.75$. Under quantum mechanics, one could theoretically achieve $p = 0.85$, corresponding to Tsirelson's bound $S \le 2 \sqrt 2$. Thus we can also obtain a p-value which is robust against violation of the independence and identical distributions assumption needed to reduce the data to multinomial counts. In this case, the Bell game p-value is the probability that a $\text{Bin}(1649, 3/4)$ distributed random variable exceeds 1356; and that turns out to equal $5 \times 10^{-13}$. (The Bell-game test was used by the experimenters in Delft and Munich; the Delft group had refined and simplified results found by the present author, 20 years earlier. See the ``supplementary material'' of Hensen et al.~(2015) \cite{hensen}.

We find this result rather exciting. Provided one has taken care of really good randomisation of measurement settings, the martingale test is hardly different from the test based on conventional calculation of $z$-values using approximate normality. The former provides, moreover, security again trends or jumps in the physical parameters of detectors. We suggest that past observations of locality violations at the manifest level of apparent correlations between Alice's setting and Bob's outcome, or vice versa, were simply manifestations of the statistical phenomenon of spurious correlations being caused by hidden confounders; the hidden confounder simply being time.

\section{Conclusion}

In a Bell test, if we are confident that there have not been shifts or jumps in the physics of the systems being studied during the course of the experiment, improved estimates of $S$ or $J$ are not difficult to obtain, and more reliable p-values can be found without much difficulty either. This can lead to big improvement of the results of experiments of the 2015 Vienna and NIST types: the Eberhard inequality is not the best test of local realism, by a long way (though not surprisingly, it is better than CHSH). Of course, many scientists will be more convinced by a very simple and very robust estimation method. Rutherford said ``if you need statistics you did the wrong experiment''. Well, some experiments do need statistics anyway. In that case one should process the data in the most efficient way possible using time honoured methods completely familiar to applied statisticians working in all fields of science. There is no excuse for the experimental physicist not to use the best tools available.

We remark that those astronomically small p-values in the Vienna and NIST experiments need to be taken with more than just a grain of salt. They are meaningless. The absolute error in the normal approximation must be huge compared to either actual or nominal (according to the normal distribution) value of these tail probabilities. In fact, we suggest that a meaningful, probably conservative, p-value is obtained from Chebyshev's inequality. A z-value of $17.5$ would correspond to a p-value of $1/17.5^2 = 0.0033$ or 3 pro mille. On the other hand, the most recent and clearly best controlled Bell experiment to date, that of Zhang et al.~(2022) \cite{zhang}, has essentially the same p-value independently of whether one uses a conventional estimation of standard error based on multinomial distributed counts and normal approximation to that distribution, or uses a martingale based statistic following the idea of the Bell game, which should insure the user against jumps and trends in the physical systems being studied over time. The close likeness of all the statistical test results suggests that there were hardly any shifts in time.

\section{References}
{\raggedright
}

\appendix
\section*{Appendix}
The appendix contains the summary statistics of the 2015 experiments of Delft, Munich (actually only published in 2017), NIST, Vienna, as well as that of Weihs et al.~(1998) at Insbruck, and Zhang et al.~(2022) in Munich again. In each case we present the raw counts and the relative frequencies, per setting combination. Rows denote Alice's outcomes, columns denote Bob's outcomes; they are conventionally taken to be $\pm 1$ in some of the experiments, and ``d'' and ``n'' standing for ``detection'' and ``nondetection'' in others. The settings have sometimes been reordered so that in cases, setting pairs (1,1), (1,2), (2,1) produce large counts (and large relative frequencies) on the diagonals, small off-diagonal; setting pair (2,2) has large numbers (and large relative frequencies) off-diagonal, small on the diagonal. In all cases, using a common notation, the sample value of the one-sided CHSH statistic $S = \rho_{11} + \rho_{12} + \rho_{22} - \rho_{22}$ exceeds $2$.

\renewcommand{\thesubsection}{\Alph{subsection}}

\subsection{Hensen et al.'s (2015) experiment at Delft \cite{hensen}}

\bigskip
\hrule

\bigskip

{\bf Raw counts}

\bigskip

\begin{minipage}{5cm}
Settings (1,1)

\begin{tabular}{rrr}
  \toprule
 & $+$ & $-$ \\ 
  \midrule
$+$ & 23 & 3 \\ 
$-$ & 4 & 23 \\ 
   \bottomrule
\end{tabular}

\end{minipage}
\begin{minipage}{5cm}
Settings (1,2)

\begin{tabular}{rrr}
  \toprule
 & + & $-$ \\ 
  \midrule
+ & 33 & 11 \\ 
  $-$ & 5 & 30 \\ 
   \bottomrule
\end{tabular}

\end{minipage}

\begin{minipage}{5cm}

\bigskip

Settings (2,1)

\begin{tabular}{rrr}
  \toprule
 & + & $-$ \\ 
  \midrule
+ & 22 & 10 \\ 
  $-$ & 6 & 24 \\ 
   \bottomrule
\end{tabular}

\end{minipage}
\begin{minipage}{5cm}

Settings (2,2)

\begin{tabular}{rrr}
  \toprule
 & + & $-$ \\ 
  \midrule
+ & 4 & 20 \\ 
 $-$ & 21 & 6 \\ 
   \bottomrule
\end{tabular}

\end{minipage}

\bigskip

{\bf Relative frequencies}

\bigskip

\begin{minipage}{5cm}
Settings (1,1)

\begin{tabular}{rrr}
  \toprule
 & + & $-$ \\ 
  \midrule
+ & 0.43 & 0.06 \\ 
  $-$ & 0.08 & 0.43 \\ 
   \bottomrule
\end{tabular}

\end{minipage}
\begin{minipage}{5cm}
Settings (1,2)

\begin{tabular}{rrr}
  \toprule
 & + & $-$ \\ 
  \midrule
+ & 0.42 & 0.14 \\ 
  $-$ & 0.06 & 0.38 \\ 
   \bottomrule
\end{tabular}

\end{minipage}

\bigskip

\begin{minipage}{5cm}

Settings (2,1)

\begin{tabular}{rrr}
  \toprule
 & + & $-$ \\ 
  \midrule
+ & 0.35 & 0.16 \\ 
  $-$ & 0.10 & 0.39 \\ 
   \bottomrule
\end{tabular}
\end{minipage}
\begin{minipage}{5cm}

Settings (2,2)

\begin{tabular}{rrr}
  \toprule
 & + & $-$ \\ 
  \midrule
+ & 0.08 & 0.39 \\ 
  $-$ & 0.41 & 0.12 \\ 
   \bottomrule
\end{tabular}

\end{minipage}

\bigskip
\hrule
\bigskip


\subsection{Rosenfeld et al.'s (2017) experiment at Munich \cite{rosenfeld}}

\bigskip 

\hrule

\bigskip

{\bf Raw counts}

\bigskip

\begin{minipage}{5cm}
Settings (1,1)

\begin{tabular}{rrr}
  \toprule
 & + & $-$ \\ 
  \midrule
+ & 16 & 4 \\ 
  $-$ & 3 & 13 \\ 
   \bottomrule
\end{tabular}

\end{minipage}
\begin{minipage}{5cm}
Settings (1,2)

\begin{tabular}{rrr}
  \toprule
 & + & $-$ \\ 
  \midrule
+ & 11 & 4 \\ 
  $-$ & 2 & 17 \\ 
   \bottomrule
\end{tabular}

\end{minipage}

\ 
\bigskip
\bigskip

\begin{minipage}{5cm}

Settings (2,1)

\begin{tabular}{rrr}
  \toprule
 & + & $-$ \\ 
  \midrule
+ & 19 & 4 \\ 
  $-$ & 3 & 16 \\ 
   \bottomrule
\end{tabular}

\end{minipage}
\begin{minipage}{5cm}

Settings (2,2)

\begin{tabular}{rrr}
  \toprule
 & + & $-$ \\ 
  \midrule
+ & 4 & 22 \\ 
  $-$ & 10 & 2 \\ 
   \bottomrule
\end{tabular}

\end{minipage}

\ 
\bigskip
\bigskip

{\bf Relative frequencies}

\bigskip

\begin{minipage}{5cm}
Settings (1,1)

\begin{tabular}{rrr}
  \toprule
 & + & $-$ \\ 
  \midrule
+ & 0.44 & 0.11 \\ 
  $-$ & 0.08 & 0.36 \\ 
   \bottomrule
\end{tabular}

\end{minipage}
\begin{minipage}{5cm}
Settings (1,2)

\begin{tabular}{rrr}
  \toprule
 & + & $-$ \\ 
  \midrule
+ & 0.32 & 0.12 \\ 
  $-$ & 0.06 & 0.50 \\ 
   \bottomrule
\end{tabular}

\end{minipage}

\ 
\bigskip
\bigskip

\begin{minipage}{5cm}

Settings (2,1)

\begin{tabular}{rrr}
  \toprule
 & + & $-$ \\ 
  \midrule
+ & 0.45 & 0.10 \\ 
  $-$ & 0.07 & 0.38 \\ 
   \bottomrule
\end{tabular}
\end{minipage}
\begin{minipage}{5cm}

Settings (2,2)

\begin{tabular}{rrr}
  \toprule
 & + & $-$ \\ 
  \midrule
+ & 0.11 & 0.58 \\ 
  $-$ & 0.26 & 0.05 \\ 
   \bottomrule
\end{tabular}

\end{minipage}

\

\bigskip

\hrule


\subsection{ Shalm et al.'s (2015) experiment at NIST (Boulder, Colorado) \cite{shalm}}

\bigskip 

\hrule

\bigskip

{\bf Raw counts}

\bigskip

\begin{minipage}{5cm}
Settings (1,1)

\begin{tabular}{rrr}
  \toprule
 & d & n \\ 
  \midrule
d & 6378 & 3282 \\ 
  n & 3189 & 43897356 \\ 
   \bottomrule
\end{tabular}

\end{minipage}
\begin{minipage}{5cm}
Settings (1,2)

\begin{tabular}{rrr}
  \toprule
 & d & n \\ 
  \midrule
d & 6794 & 2821 \\ 
  n & 23243 & 43276943 \\ 
   \bottomrule
\end{tabular}

\end{minipage}

\ 
\bigskip
\bigskip

\begin{minipage}{5cm}

Settings (2,1)

\begin{tabular}{rrr}
  \toprule
 & d & n \\ 
  \midrule
d & 6486 & 21334 \\ 
  n & 2843 & 43338281 \\ 
   \bottomrule
\end{tabular}

\end{minipage}
\begin{minipage}{5cm}

Settings (2,2)

\begin{tabular}{rrr}
  \toprule
 & d & n \\ 
  \midrule
d & 106 & 27539 \\ 
  n & 30040 & 42502788 \\ 
   \bottomrule
\end{tabular}

\end{minipage}

\ 
\bigskip
\bigskip

{\bf Relative frequencies$\times 10^6$}

The bottom right cell in all four $2\times 2$ tables is close to $10^6$, the difference is negligeable.

\bigskip

\begin{minipage}{5cm}
Settings (1,1)

\begin{tabular}{rrr}
  \toprule
 & d & n \\ 
  \midrule
d & 145.25 & 74.74 \\ 
  n & 72.63 & 999707.38 \\ 
   \bottomrule
\end{tabular}

\end{minipage}
\begin{minipage}{5cm}
Settings (1,2)

\begin{tabular}{rrr}
  \toprule
 & d & n \\ 
  \midrule
d & 156.87 & 65.14 \\ 
  n & 536.67 & 999241.33 \\ 
   \bottomrule
\end{tabular}

\end{minipage}

\ 
\bigskip
\bigskip

\begin{minipage}{5cm}

Settings (2,1)

\begin{tabular}{rrr}
  \toprule
 & d & n \\ 
  \midrule
d & 149.55 & 491.92 \\ 
  n & 65.55 & 999292.97 \\ 
   \bottomrule
\end{tabular}
\end{minipage}
\begin{minipage}{5cm}

Settings (2,2)

\begin{tabular}{rrr}
  \toprule
 & d & n \\ 
  \midrule
d & 2.49 & 647.06 \\ 
  n & 705.82 & 998644.63 \\ 
   \bottomrule
\end{tabular}

\end{minipage}

\ 
\bigskip

\hrule


\subsection{ Giustina et al.'s (2015) experiment at Vienna \cite{giustina}}

\bigskip 

\hrule

\bigskip

{\bf Raw counts}

\bigskip

\begin{minipage}{5cm}
Settings (1,1)

\begin{tabular}{rrr}
  \toprule
 & d & n \\ 
  \midrule
d & 141439 & 73391 \\ 
  n & 76224 & 875392736 \\ 
   \bottomrule
\end{tabular}

\end{minipage}
\begin{minipage}{5cm}
Settings (1,2)

\begin{tabular}{rrr}
  \toprule
 & d & n \\ 
  \midrule
d & 146831 & 67941 \\ 
  n & 326768 & 874976534 \\ 
   \bottomrule
\end{tabular}

\end{minipage}

\ 
\bigskip
\bigskip

\begin{minipage}{5cm}

Settings (2,1)

\begin{tabular}{rrr}
  \toprule
 & d & n \\ 
  \midrule
d & 158338 & 425067 \\ 
  n & 58742 & 875239860 \\ 
   \bottomrule
\end{tabular}

\end{minipage}
\begin{minipage}{5cm}

Settings (2,2)

\begin{tabular}{rrr}
  \toprule
 & d & n \\ 
  \midrule
d & 8392 & 576445 \\ 
  n & 463985 & 874651457 \\ 
   \bottomrule
\end{tabular}

\end{minipage}

\ 
\bigskip
\bigskip

{\bf Relative frequencies$\times 10^6$}

The bottom right cell in all four $2\times 2$ tables is close to $10^6$, the difference is negligeable.

\bigskip

\begin{minipage}{5cm}
Settings (1,1)

\begin{tabular}{rrr}
  \toprule
 & d & n \\ 
  \midrule
d & 161.52 & 83.81 \\ 
  n & 87.05 & 999667.63 \\ 
   \bottomrule
\end{tabular}

\end{minipage}
\begin{minipage}{5cm}
Settings (1,2)

\begin{tabular}{rrr}
  \toprule
 & d & n \\ 
  \midrule
d & 167.71 & 77.60 \\ 
  n & 373.23 & 999381.46 \\ 
   \bottomrule
\end{tabular}

\end{minipage}

\ 
\bigskip
\bigskip

\begin{minipage}{5cm}

Settings (2,1)

\begin{tabular}{rrr}
  \toprule
 & d & n \\ 
  \midrule
d & 180.78 & 485.30 \\ 
  n & 67.07 & 999266.86 \\ 
   \bottomrule
\end{tabular}
\end{minipage}
\begin{minipage}{5cm}

Settings (2,2)

\begin{tabular}{rrr}
  \toprule
 & d & n \\ 
  \midrule
d & 9.58 & 658.27 \\ 
  n & 529.84 & 998802.30 \\ 
   \bottomrule
\end{tabular}

\end{minipage}

\ 
\bigskip

\hrule


\subsection{ Weihs et al.~(1998) Innsbruck experiment \cite{weihs}}

\bigskip 

\hrule

\bigskip

{\bf Raw counts}

\bigskip

\begin{minipage}{5cm}
Settings (1,1)

\begin{tabular}{rrr}
  \toprule
 & $+$ & $-$ \\ 
  \midrule
$+$ & 1683 & 418 \\ 
$-$ & 361 & 1578 \\ 
   \bottomrule
\end{tabular}

\end{minipage}
\begin{minipage}{5cm}
Settings (1,2)

\begin{tabular}{rrr}
  \toprule
 & + & $-$ \\ 
  \midrule
+ & 1100 & 269 \\ 
  $-$ & 156 & 1386 \\ 
   \bottomrule
\end{tabular}

\end{minipage}

\ 
\bigskip
\bigskip

\begin{minipage}{5cm}

Settings (2,1)

\begin{tabular}{rrr}
  \toprule
 & + & $-$ \\ 
  \midrule
+ & 1728 & 313 \\ 
  $-$ & 351 & 1978 \\ 
   \bottomrule
\end{tabular}

\end{minipage}
\begin{minipage}{5cm}

Settings (2,2)

\begin{tabular}{rrr}
  \toprule
 & + & $-$ \\ 
  \midrule
+ & 179 & 1636 \\ 
 $-$ & 1143 & 294 \\ 
   \bottomrule
\end{tabular}

\end{minipage}

\ 
\bigskip
\bigskip

{\bf Relative frequencies}

\bigskip

\begin{minipage}{5cm}
Settings (1,1)

\begin{tabular}{rrr}
  \toprule
 & + & $-$ \\ 
  \midrule
+ & 0.42 & 0.10 \\ 
  $-$ & 0.09 & 0.39 \\ 
   \bottomrule
\end{tabular}

\end{minipage}
\begin{minipage}{5cm}
Settings (1,2)

\begin{tabular}{rrr}
  \toprule
 & + & $-$ \\ 
  \midrule
+ & 0.38 & 0.09 \\ 
  $-$ & 0.05 & 0.48 \\ 
   \bottomrule
\end{tabular}

\end{minipage}

\ 
\bigskip
\bigskip

\begin{minipage}{5cm}

Settings (2,1)

\begin{tabular}{rrr}
  \toprule
 & + & $-$ \\ 
  \midrule
+ & 0.40 & 0.07 \\ 
  $-$ & 0.08 & 0.45 \\ 
   \bottomrule
\end{tabular}
\end{minipage}
\begin{minipage}{5cm}

Settings (2,2)

\begin{tabular}{rrr}
  \toprule
 & + & $-$ \\ 
  \midrule
+ & 0.06 & 0.50 \\ 
  $-$ & 0.35 & 0.09 \\ 
   \bottomrule
\end{tabular}

\end{minipage}

\ 

\bigskip

\hrule

\subsection{ Zhang et al.'s (2022) experiment at Munich \cite{zhang}}

\bigskip 

\hrule

\bigskip

{\bf Raw counts}

\bigskip

\begin{minipage}{5cm}
Settings (1,1)

\begin{tabular}{rrr}
  \toprule
 & $+$ & $-$ \\ 
  \midrule
$+$ & 178 & 44 \\ 
$-$ & 29 & 183 \\ 
   \bottomrule
\end{tabular}

\end{minipage}
\begin{minipage}{5cm}
Settings (1,2)

\begin{tabular}{rrr}
  \toprule
 & + & $-$ \\ 
  \midrule
+ & 199 & 36 \\ 
  $-$ & 28 & 160 \\ 
   \bottomrule
\end{tabular}

\end{minipage}

\ 
\bigskip
\bigskip

\begin{minipage}{5cm}

Settings (2,1)

\begin{tabular}{rrr}
  \toprule
 & + & $-$ \\ 
  \midrule
+ & 160 & 47 \\ 
  $-$ & 31 & 15 1 \\ 
   \bottomrule
\end{tabular}

\end{minipage}
\begin{minipage}{5cm}

Settings (2,2)

\begin{tabular}{rrr}
  \toprule
 & + & $-$ \\ 
  \midrule
+ & 38 & 160 \\ 
 $-$ & 166 & 39 \\ 
   \bottomrule
\end{tabular}

\end{minipage}

\ 
\bigskip
\bigskip

{\bf Relative frequencies}

\bigskip

\begin{minipage}{5cm}
Settings (1,1)

\begin{tabular}{rrr}
  \toprule
 & + & $-$ \\ 
  \midrule
+ & 0.41 & 0.10 \\ 
  $-$ & 0.07 & 0.42 \\ 
   \bottomrule
\end{tabular}

\end{minipage}
\begin{minipage}{5cm}
Settings (1,2)

\begin{tabular}{rrr}
  \toprule
 & + & $-$ \\ 
  \midrule
+ & 0.47 & 0.09 \\ 
  $-$ & 0.07 & 0.38 \\ 
   \bottomrule
\end{tabular}

\end{minipage}

\ 
\bigskip
\bigskip

\begin{minipage}{5cm}

Settings (2,1)

\begin{tabular}{rrr}
  \toprule
 & + & $-$ \\ 
  \midrule
+ & 0.41 & 0.12 \\ 
  $-$ & 0.08 & 0.39 \\ 
   \bottomrule
\end{tabular}
\end{minipage}
\begin{minipage}{5cm}

Settings (2,2)

\begin{tabular}{rrr}
  \toprule
 & + & $-$ \\ 
  \midrule
+ & 0.09 & 0.40 \\ 
  $-$ & 0.41 & 0.10 \\ 
   \bottomrule
\end{tabular}

\end{minipage}

\ 

\bigskip

\hrule

\end{document}